\documentclass[twocolumn]{aa}
\usepackage{graphicx}
\usepackage{txfonts}
\begin{document}

   \title{ Deuterium depletion and magnesium enhancement in the local disc }

   \subtitle{}

   \author{P. Gnaci\'nski }
   \authorrunning{P. Gnaci\'nski}
          

   \institute{Institute of Theoretical Physics and Astrophysics,
              University of Gda\'nsk,
              ul. Wita Stwosza 57, 80-952 Gda\'nsk, Poland\\
              \email{pg@iftia.univ.gda.pl}
         }

   \date{}

   \abstract
   {}
   { The local disc deuter is known to be depleted in comparison to the local bubble. It is not clear which process has caused this depletion. }
   { We have used Hubble Space Telescope (HST) spectra to obtain column densities of Si,        Mg and Fe elements. We have compared normalized column densities of this elements in       the directions with high and low deuterium abundances. }
   { We show, that the same lines of sight that are depleted in deuter, are enhanced in magnesium. Heavier elements - Si and Fe do not show any difference in the abundance between the local disc and the local bubble. This observation implicates that astration is responsible for both deuter depletion and magnesium enhancement.
   }
   {} 
   

   \maketitle

\section{Introduction}

  Deuter is one of the most interesting isotopes in astrophysics. Deuter was produced in
the Big Bang nucleosynthesis and plays an important role in testing evolution models of the Universe. The deuterium abundance in the interstellar medium tells us about stellar 
processing of the matter.

  The gas phase deuterium abundance changes with the distance or with the hydrogen 
column density (\cite{Wood}). It was not clear, what process is responsible for the deuterium depletion. Several processes were considered in this context, like astration or selective binding of deuterium in molecules or dust grains. The aim of this paper is to check which process is responsible for the deuterium depletion.
  
\section{Column densities}

  We have selected our target stars from the compilation by \cite{Wood}. From their table 4 we have chosen stars with HST high resolution observations. Both GHRS and STIS spectra
were considered. We have selected two groups of stars: one with $19.2<\log N(HI)$ and (D/H)$\approx 1.5\cdot10^{-5}$, and the second with low D/H ratio (D/H$<1\cdot10^{-5}$).
The HST spectra used in this paper are collected in Table \ref{widma}.

\begin{table}
\scriptsize
\begin{tabular}{lccc}
\hline \hline
Star       & Mg II  &  Si II   & Fe II  \\
\hline
HD 22049   & o55p01010              & ---          & ---           \\
HD 61421   & z17x0404               & ---          & ---           \\
HD 62509   & z2si0404               & ---          & ---           \\
HD 34029   & z18v030a,z0jr010x      & ---          & ---           \\
HD 432     & z2si0105               & ---          & ---           \\
HD 11443   & z2si0205               & ---          & ---           \\
HD 22468   & z1gu0503, z1gv0[28c]06 & ---          & ---           \\
HD 4128    & z2dc0208               & ---          & ---           \\
HZ 43      & z2r50205               & ---          & ---           \\
HD 62044   & z2si0[38]04            & ---          & ---           \\
G191-B2B   & z14z010o,o6hb300[89abcd]0  & ---      & ---           \\
HD 36486   & z2xu010a               & z185020[89a] & z185020[nop]  \\
HD 37128   & z1bw040u               & z1bw030[ijk] & z1bw030[fgh]  \\
HD 38666   & z2az010o               & z2d40211     & z2d4030u      \\
HD 158926  & z1bw080lmn             & z1bw070[ijk] & z1bw070[fgh]  \\
HD 195965  & o6bg01010              & o6bg01030    & ---           \\
\hline
\end{tabular}
\normalsize
\caption{ HST spectra used to derive column densities of the Mg, Si and Fe elements.
  \label{widma}
  }
\end{table}

  The column densities were derived using the profile fitting technique. The absorption
lines were fitted by Voigt profiles, except the Mg II 1240 \AA\ doublet, which was fitted by a Gauss function. The cloud velocities, Doppler broadening parameters and column densities for multiple absorption components were simultaneously fitted to the observed spectrum. Both lines of magnesium doublet (at 2800 \AA\ or 1200 \AA) were also fitted simultaneously. A convolution with a point spread function was performed. The wavelengths, oscillator strengths and natural damping constants were adopted from \cite{Morton}.
The derived column densities are presented in Table \ref{CD}.

\begin{figure*}
   \centering
   \includegraphics[width=160mm]{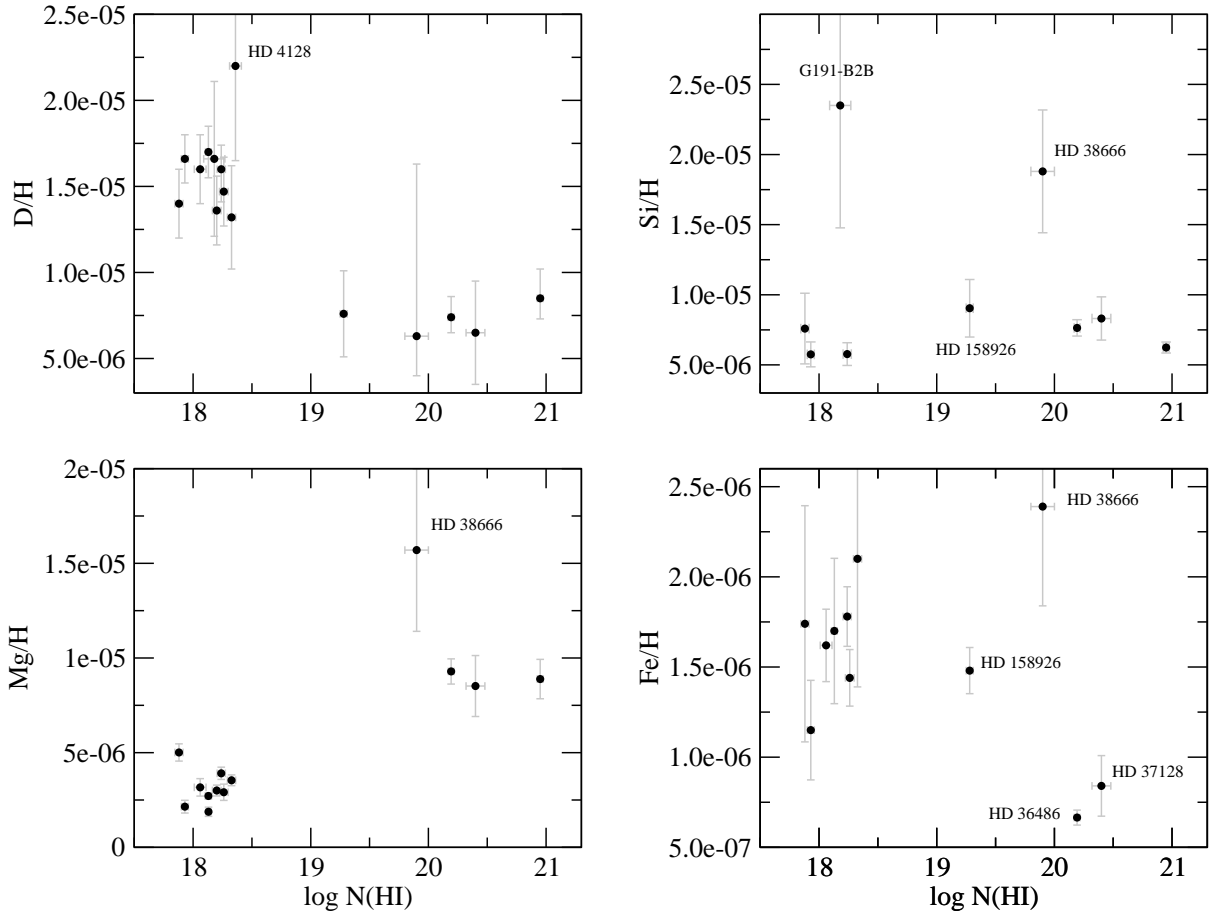}
      \caption{ 
        Element abundances vs. hydrogen column density.
              }
      \label{abundances}
\end{figure*}

\begin{table*}
\begin{tabular}{llllll}
\hline \hline
Star      & Other name     & log N(HI)             & log N(SiII)       & log N(MgII)       & log N(FeII)   \\
          &                & $[\log($cm$^{-2})]$     & $[\log($cm$^{-2})]$ & $[\log($cm$^{-2})]$ & $[\log($cm$^{-2})]$ \\
\hline
          &                & ref. (1)              &  ref. (2)         & ref. (4)          & ref. (3)    \\
HD 22049  & $\epsilon$ Eri & 17.88 \ \ $\pm$ 0.035 & 12.76 $\pm$ 0.14  & 12.58 $\pm$ 0.02  & 12.12 $\pm$ 0.16 \\
HD 61421  & Procyon        & 18.06 \ \ $\pm$ 0.05  &                   & 12.56 $\pm$ 0.04  & 12.27 $\pm$ 0.02 \\
HD 62509  & $\beta$ Gem    & 18.261 $\pm$ 0.037    &                   & 12.72 $\pm$ 0.05  & 12.42 $\pm$ 0.03 \\
HD 34029  & Capella        & 18.239 $\pm$ 0.035    & 13.00 $\pm$ 0.05  & 12.83 $\pm$ 0.02  & 12.49 $\pm$ 0.02 \\
HD 432    & $\beta$ Cas    & 18.13 \ \ $\pm$ 0.025 &                   & 12.56 $\pm$ 0.01  & 12.36 $\pm$ 0.10 \\
HD 11443  & $\alpha$ Tri   & 18.327 $\pm$ 0.035    &                   & 12.876$\pm$ 0.002 & 12.65 $\pm$ 0.14 \\
HD 22468  & HR 1099        & 18.131 $\pm$ 0.02     &                   & 12.40 $\pm$ 0.05  &                  \\
HD 4128   & $\beta$ Cet    & 18.36 \ \ $\pm$ 0.05  & 14.17 $\pm$ 0.34  & saturated         &                  \\
HZ 43     &                & 17.93 \ \ $\pm$ 0.03  & 12.69 $\pm$ 0.06  & 12.26 $\pm$ 0.06  & 11.99 $\pm$ 0.10 \\
HD 62044  & $\sigma$ Gem   & 18.201 $\pm$ 0.037    &                   & 12.68 $\pm$ 0.02  &                  \\ 
G191-B2B  & BD +52 913     & 18.18 \ \ $\pm$ 0.09  & 13.55 $\pm$ 0.13  & saturated         & 13.08            \\ 
\hline
          &                & ref. (1)              & ref. (4)          & ref. (4)          & ref. (4) \\
HD 158926 & $\lambda$ Sco  & 19.28 \ \ $\pm$ 0.03  & 14.24 $\pm$ 0.09  & saturated         & 13.45 $\pm$ 0.02 \\
HD 36486  & $\delta$ Ori   & 20.193 $\pm$ 0.025    & 15.08 $\pm$ 0.02  & 15.16 $\pm$ 0.02  & 14.02 $\pm$ 0.01 \\
HD 38666  & $\mu$ Col      & 19.9 \ \ \ \ $\pm$ 0.1 & 15.17 $\pm$ 0.01 & 15.10 $\pm$ 0.06  & 14.278 $\pm$ 0.004 \\
HD 37128  & $\epsilon$ Ori & 20.40 \ \ $\pm$ 0.08  & 15.32 $\pm$ 0.01  & 15.33  $\pm$ 0.02  & 14.32 $\pm$ 0.03 \\
HD 195965 & BD +47 3136    & 20.95 \ \ $\pm$ 0.025 & 15.74 $\pm$ 0.01  & 15.90 $\pm$ 0.04  &                         \\
\hline
\end{tabular}

\caption{ 
Column densities for analysed sight lines. References: (1)-\cite{Wood} ; (2)-\cite{Redfield2004} ; (3)-\cite{Redfield2002} ; (4)-this paper. \label{CD}
  }
\end{table*}

  The Si, Mg and Fe column densities normalized by the hydrogen column density are presented in Table \ref{CD/H}. The direction to the star HD 4128 ($\beta$ Cet) presents
an unexpected high Si/H ratio. The Si/H$=6.5\cdot10^{-5}$ is much higher than the solar
system abundance (Si/H)$_\odot=3.63\cdot10^{-5}$. Also the D/H$=2.2\cdot10^{-5}$ is 
higher than for other sightlines. Therefore the column densities for the star HD 4128 were excluded from further consideration. For the white dwarf G191-B2B the Fe/H$=7.95\cdot10^{-6}$ value is 4 times
higher than in all other directions. This value was also excluded from further
calculations.

\begin{table*}
\begin{tabular}{llrllll}
\hline \hline
Star      & Other name     &  \multicolumn{2}{c}{D/H}                                        &  \multicolumn{1}{c}{Si/H}                & \multicolumn{1}{c}{Mg/H}                & \multicolumn{1}{c}{Fe/H}  \\
\hline
HD 22049  & $\epsilon$ Eri &  $1.4\cdot10^{-5}$  & $ \pm   2\cdot10^{-6} $                   & $7.6\cdot10^{-6} \pm   2.5\cdot10^{-6}$  & $5.0\cdot10^{-6} \pm  4.6\cdot10^{-7} $ & $ 1.7\cdot10^{-6} \pm 6.6\cdot10^{-7} $ \\
HD 61421  & Procyon        &  $1.6\cdot10^{-5}$  & $ \pm   2\cdot10^{-6} $                   &                                          & $3.2\cdot10^{-6} \pm  4.6\cdot10^{-7} $ & $ 1.6\cdot10^{-6} \pm 2.0\cdot10^{-7} $ \\
HD 62509  & $\beta$ Gem    &  $1.47\cdot10^{-5}$ & $ \pm   2\cdot10^{-6} $                   &                                          & $2.9\cdot10^{-6} \pm  4.3\cdot10^{-7} $ & $ 1.4\cdot10^{-6} \pm 1.6\cdot10^{-7} $ \\
HD 34029  & Capella        &  $1.6\cdot10^{-5}$  & $^{+1.4\cdot10^{-6}} _{-1.9\cdot10^{-6}}$ & $5.8\cdot10^{-6} \pm   8.1\cdot10^{-7}$  & $3.9\cdot10^{-6} \pm  3.2\cdot10^{-7} $ & $ 1.8\cdot10^{-6} \pm 1.7\cdot10^{-7} $ \\
HD 432    & $\beta$ Cas    &  $1.7\cdot10^{-5}$  & $ \pm   1.5\cdot10^{-6} $                 &                                          & $2.7\cdot10^{-6} \pm  1.6\cdot10^{-7} $ & $ 1.7\cdot10^{-6} \pm 4.0\cdot10^{-7} $ \\
HD 11443  & $\alpha$ Tri   &  $1.32\cdot10^{-5}$ & $ \pm   3\cdot10^{-6}   $                 &                                          & $3.5\cdot10^{-6} \pm  2.9\cdot10^{-7} $ & $ 2.1\cdot10^{-6} \pm 7.1\cdot10^{-7} $ \\
HD 22468  & HR 1099        &  $1.46\cdot10^{-5}$ & $ \pm   9\cdot10^{-7}   $                 &                                          & $1.9\cdot10^{-6} \pm  2.4\cdot10^{-7} $ &   \\
HD 4128   & $\beta$ Cet    &  $\it{2.2\cdot10^{-5}}$ & $ \it{\pm 5.5\cdot10^{-6}}$           & $\it{6.5\cdot10^{-5} \pm 5.1\cdot10^{-5}}$ &                                       &   \\
HZ 43     &                &  $1.66\cdot10^{-5}$ & $ \pm   1.4\cdot10^{-6} $                 & $5.8\cdot10^{-6} \pm   8.9\cdot10^{-7}$  & $2.2\cdot10^{-6} \pm  3.4\cdot10^{-6} $ & $ 1.1\cdot10^{-6} \pm 2.8\cdot10^{-7} $ \\
HD 62044  & $\sigma$ Gem   &  $1.36\cdot10^{-5}$ & $ \pm   2\cdot10^{-6}   $                 &                                          & $3.0\cdot10^{-6} \pm  2.8\cdot10^{-7} $ &   \\
G191-B2B  & BD +52 913     &  $1.66\cdot10^{-5}$ & $ \pm   4.5\cdot10^{-6} $                 & $2.4\cdot10^{-5} \pm   8.7\cdot10^{-6}$  &                                         & $ \it{7.9\cdot10^{-6} \pm 1.6\cdot10^{-6}} $ \\
\hline
HD 158926 & $\lambda$ Sco  &  $7.6\cdot10^{-6}$  & $  \pm   2.5\cdot10^{-6}  $               & $9.0\cdot10^{-6} \pm   2.1\cdot10^{-6}$  &                                         & $ 1.5\cdot10^{-6} \pm 1.3\cdot10^{-7} $ \\
HD 36486  & $\delta$ Ori   &  $7.4\cdot10^{-6}$  & $^{+1.2\cdot10^{-6}} _{-9\cdot10^{-7}} $  & $7.6\cdot10^{-6} \pm   5.8\cdot10^{-7}$  & $9.3\cdot10^{-6} \pm  6.6\cdot10^{-7} $ & $ 6.6\cdot10^{-7} \pm 4.1\cdot10^{-8} $ \\
HD 38666  & $\mu$ Col      &  $6.3\cdot10^{-6}$  & $^{+1\cdot10^{-5}} _{-2.3\cdot10^{-6}} $  & $1.9\cdot10^{-5} \pm   4.4\cdot10^{-6}$  & $1.6\cdot10^{-5} \pm  4.3\cdot10^{-6} $ & $ 2.4\cdot10^{-6} \pm 5.5\cdot10^{-7} $ \\
HD 37128  & $\epsilon$ Ori &  $6.5\cdot10^{-6}$  & $ \pm   3\cdot10^{-6}                   $ & $8.3\cdot10^{-6} \pm   1.5\cdot10^{-6}$  & $8.5\cdot10^{-6} \pm  1.6\cdot10^{-6} $ & $ 8.4\cdot10^{-7} \pm 1.7\cdot10^{-7} $ \\
HD 195965 & BD +47 3136    &  $8.5\cdot10^{-6}$  & $^{+1.7\cdot10^{-6}} _{-1.2\cdot10^{-6}}$ & $6.2\cdot10^{-6} \pm   3.9\cdot10^{-7}$  & $8.9\cdot10^{-6} \pm  1.0\cdot10^{-6} $ &   \\
\hline
\multicolumn{2}{c}{Avarage (D/H$>1\cdot10^{-5}$)} &  $1.52\cdot10^{-5}$ & $\pm 1.4\cdot10^{-6}$ & $1.07\cdot10^{-5} \pm   8.6\cdot10^{-6}$ & $3.14\cdot10^{-6} \pm 9.4\cdot10^{-7} $ & $ 1.65\cdot10^{-6} \pm 3.0\cdot10^{-7} $ \\
 
\multicolumn{2}{c}{Avarage (D/H$\le1\cdot10^{-5}$)} &  $7.26\cdot10^{-6}$ & $\pm 8.9\cdot10^{-7}$ & $1.00\cdot10^{-5} \pm   5.0\cdot10^{-6}$ & $1.06\cdot10^{-5} \pm 3.42\cdot10^{-6} $ & $ 1.34\cdot10^{-6} \pm 7.8\cdot10^{-7} $ \\
 
\multicolumn{2}{c}{Student's t-distribution}  &  3.84               &                         & 0.07                                     & -2.88                                   &   0.44  \\
\multicolumn{2}{c}{significance level}        &  0.002              &                         & 0.95                                     & 0.015                                    &  0.67   \\
\hline

\end{tabular}
\caption{ 
  Normalized elements abundances. The D/H ratio was taken from the compilation by \cite{Wood}. Averages for sight lines with the D/H$>1\cdot10^{-5}$ and for lower D/H
ratios are shown. The significance level of conformance between these two averages is shown in the last row. Three values withdrawn from the calculation are shown in italics (HD 4128 and G191-B2B).  \label{CD/H}
 }
\end{table*}


\section{Results}
 
  The abundances of the D, Si , Mg and Fe elements are presented on Fig. \ref{abundances}.
For sigh lines with $\log H>19.2$ and low D/H ratio an enhancement of magnesium is
clearly seen. For the Si and Fe elements there are no systematic differences in the abundance.
 
  To test the conformance of elements abundances in sight lines of different D/H ratio we
have used the Student's variable. The Student's variable $t$ describing conformance of two average
values $\bar{x}$ and $\bar{y}$ was calculated with the formula:
  \begin{equation}
    t=\frac{\bar{x}-\bar{y}}{\sqrt{\frac{n+m}{n+m-2}}\sqrt{\frac{n-1}{m}S^2_{\bar{x}}+
    \frac{m-1}{n}S^2_{\bar{y}} }}
  \end{equation}
The symbol $n$ denotes the number of measurements in $\bar{x}$, and $m$ represents the number of measurements taken into account for calculating $\bar{y}$. $S_{\bar{x}}$ and  $S_{\bar{y}}$ are standard deviations for a sample. 
 
  The conformance of element abundances in sight lines with high D/H ratio (D/H$>1\cdot10^{-5}$) and low D/H (D/H$<1\cdot10^{-5}$) was tested. The results -- the
t-Student's values and significance levels are presented in Table \ref{CD/H}.
The deuterium (significance level 0.002) and magnesium (significance level 0.015)
elements abundance is different in the two samples of stars. 
The silicon abundances in regions with high and low D/H ratio agree at significance level 0.95. No clear conclusion can be stated about the Fe abundance (significance level 0.67), but from the
last panel on Fig. \ref{abundances} one can notice that the Fe abundance is not enhanced in 
the directions with large hydrogen column densities.
  
  The difference in the magnesium abundance could potentially be assigned to the inaccuracy of
oscillator strengths, since the magnesium column density for the four stars with the large Mg/H was measured from the 1240 \AA\ doublet. \cite{Morton} cites five oscillator strength values for the Mg II 1240 \AA\ doublet, and the value preferred by \cite{Morton} and used 
in this paper is the largest one. The lowest oscillator strength value is less by 15\%
from that one used in this paper. This f-value would lead to the magnesium column density
even grater than ours (by 15\%)! \cite{Fitzpatrick} has empirically determined the f-values for the Mg II 1240 \AA\ doublet from astrophysical observations. His f-value is within one sigma from f-value used here. 
The Mg/H ratio in various directions differs by a factor $\sim 3$. 
The possibility, that the oscillator strength is inaccurate by such a large factor seems
to be unlikely, although can not be completely ruled out.

  Recently \cite{Prochaska} have considered the titanium depletion in comparison to the D/H
ratio. They have found a correlation between Ti/H and D/H, yielding a conclusion that 
the D/H scatter is caused by differential depletion of deuterium onto dust grains. On our Fig. \ref{abundances} two stars also shows Fe depletion in directions with low D/H. 
The conclusion, that deuterium is depleted on dust grains, is in opposition to the result shown in this paper. Clearly, more investigations are needed to solve
the problem of deuterium depletion.
  
\section{Conclusions}

  The sight lines with deuterium depletion show also enhancement of magnesium.
Such connection indicates that the deuterium depletion is caused by astration.
The magnesium element is produced in stellar nucleosynthesis, which also destroys deuterium. The stellar nucleosynthesis did not reach the heavier elements -- like Si and Fe, because these elements abundances are the same in the directions with low and high D/H. 
  It would be very interesting to check, if the elements lighter than Mg, like
C, N and O, are also more abundant in the directions with deuterium depletion.

\begin{acknowledgements}
   This publication is based on observations made with the NASA/ESA Hubble Space Telescope, obtained from the data archive at the Space Telescope Science Institute. STScI is operated by the Association of Universities for Research in Astronomy, Inc. under NASA contract NAS 5-26555.

\end{acknowledgements}

\end{document}